\documentclass[10pt,aps,prd,preprint,onecolumn]{revtex4}

\usepackage{amssymb}
\usepackage{amsmath}
\usepackage{graphics}
\usepackage{latexsym}
\usepackage{array}
\usepackage[activeacute,english]{babel}
\usepackage[latin1]{inputenc}
\usepackage[dvips]{graphicx}
\usepackage{color}
\usepackage{epstopdf}
\usepackage{amssymb}
\usepackage{slashed}
\usepackage{mathrsfs}
\usepackage{amsmath}
\usepackage{wasysym}

\newcommand{\bn}{\begin{eqnarray}}
\newcommand{\en}{\end{eqnarray}}
\newcommand{\be}{\begin{equation}}
\newcommand{\ee}{\end{equation}}

\newcommand{\bc}{\begin{center}}
\newcommand{\ec}{\end{center}}

\newcommand{\ket}[1]{\ensuremath{\left|#1\right\rangle}}
\newcommand{\bra}[1]{\ensuremath{\left\langle#1\right|}}

\newcommand{\parc}[2]{\frac{\partial#1}{\partial #2}}

\newcommand{\abs}[1]{\left\vert#1\right\vert}
\newcommand{\crea}[2]{\hat{#1}^{\dag}_{#2}}
\newcommand{\des}[2]{\hat{#1}_{#2}}

\newcommand{\corc}[1]{\left(#1\right)}

\begin{document}

\title{\Large \bf Spreading of wave packets for neutrino oscillations in vacuum}

\author{Y. F. P\'{e}rez}
\email{yfperezg@fmail.if.usp.br}

\affiliation{Departamento de F\'{\i}sica Matem\'{a}tica, Instituto
de F\'{\i}sica, Universidade de S\~{a}o Paulo, S\~{a}o Paulo,
Brasil}

\author{C. J. Quimbay\footnote{Associate researcher of Centro
Internacional de F\'{\i}sica, Bogot\'{a} D.C., Colombia.}}
\email{cjquimbayh@unal.edu.co}

\affiliation{Departamento de F\'{\i}sica, Universidad Nacional de Colombia.\\
Ciudad Universitaria, Bogot\'{a} D.C., Colombia.}

\date{\today}

\begin{abstract}
The effects originated in dispersion with time on spreading of
wave packets for the time-integrated two-flavor neutrino
oscillation probabilities in vacuum are studied in the context of
a field theory treatment. The neutrino flavor states are written
as superpositions of neutrino mass eigenstates which are described
by localized wave packets. This study is performed for the limit
of nearly degenerate masses and considering an expansion of the
energy until third order in the momentum. We obtain that the
time-integrated neutrino oscillation probabilities are suppressed
by a factor $1/L^2$ for the transversal and longitudinal
dispersion regimes, where $L$ is the distance between the neutrino
source and the detector.

\end{abstract}


\maketitle

\section{Introduction}\label{sec:01}

Knowing the nature of neutrino fields is an open problem in
particle physics \cite{CK93,RM02,GK07}. This problem might be
solved, experimentally, establishing if neutrinos are: (i)
Majorana fermions; (ii) Dirac fermions. For the first case,
neutrino and anti-neutrino are the same particle being described
by two-component spinorial fields called Majorana fields
\cite{CK93,RM02,GK07}. For the second case, neutrinos and
anti-neutrinos are described by four-component spinorial fields
called Dirac fields \cite{CK93,RM02,GK07}. However, the
theoretical description of neutrino oscillations leads to the same
results independently if neutrinos are Majorana or Dirac fermions.
This argument was established many years ago using the plane wave
formalism of quantum mechanics \cite{Bilenky1980}. Moreover, the
validity of this argument in the context of a quantum field theory
treatment can also been proved using the plane wave formalism
\cite{PQ1}.

The standard neutrino oscillation probabilities using the plane
wave formalism have been obtained in the context of several
quantum mechanics treatments (for instance, see
\cite{Ponteco1,Kayser,Giunti1991,Rich,Zralec,Giunti2001}). On the
other hand, in the context of different quantum field theory
treatments, neutrino oscillations in vacuum have been extensively
studied describing neutrinos by Dirac fields
\cite{Giunti1993,Giunti1998-1,Giunti1998-2,Beuthe2002,Giunti2002,Giunti2003,Beuthe2003,
Giunti2004,Giunti2007,Bilenky2011,Bilenky2011-2,Bernardini2004,Bernardini2005}.
In particular, the effects of the spreading of the wave packets on
neutrino oscillation probabilities have been widely investigated
for the case of considering an expansion of the energy until
second order in the momentum
\cite{Giunti1993,Giunti1998-1,Giunti1998-2,Beuthe2002,Giunti2002,Giunti2003,Beuthe2003,
Giunti2004,Giunti2007,Bilenky2011,Bilenky2011-2}. For this case,
the standard neutrino oscillation probabilities using the wave
packet formalism are written in terms of the oscillation and
coherence lengths
\cite{Giunti1993,Giunti1998-1,Giunti1998-2,Beuthe2002,Giunti2002,Giunti2003,Beuthe2003,
Giunti2004,Giunti2007,Bilenky2011,Bilenky2011-2}. Moreover, some
aspects of the time effects on spreading of wave packets for
neutrino oscillation probabilities have been also studied
\cite{Beuthe2002,Beuthe2003}. Nevertheless, for the case of
considering an expansion of the energy until third order in the
momentum, the study of the effects originated in dispersion with
time on spreading of wave packets for neutrino oscillation
probabilities have not been studied until now.

The main goal of this work is to study the effects originated by
dispersion in time on spreading of wave packets for the
time-integrated two-flavor neutrino oscillations in vacuum. To do
it, we perform an expansion of the energy until third order and we
consider the limit of nearly degenerate neutrino masses. The
time-integrated two-flavor neutrino oscillation probabilities are
calculated in the context of a wave packet extension of the
quantum field theory treatment that we previously developed for
the case in which neutrinos were considered as Majorana fermions
and neutrino mass eigenstates were described by plane waves
\cite{PQ1}. In the present treatment, the neutrino flavor states
are considered as superpositions of neutrino mass eigenstates
described by localized wave packets. By methodological reasons,
the effects of the spreading of the wave packets are studied for
two cases: (i) Considering the expansion of the energy until
second order in the momentum that leads to the standard
time-integrated neutrino oscillation probabilities
\cite{Giunti1993,Giunti1998-1,Giunti1998-2,Beuthe2002,Giunti2002,Giunti2003,Beuthe2003,
Giunti2004,Giunti2007,Bilenky2011,Bilenky2011-2}; (ii) considering
the expansion of the energy until third order in the momentum
that, leads to a suppression of the time-integrated neutrino
oscillation probabilities by a factor $1/L^2$ for transversal and
longitudinal dispersion regimes, where $L$ is the distance between
the neutrino source and the detector. This suppression factor is a
new result in the context of neutrino oscillation probabilities
and is in agreement with the one obtained by Naumov, whom has
recently demonstrated for a theory of wave packets that the
integral over time of both the flux and probability densities are
proportional to a factor $1/L^2$, considering the energy expanded
until third order in the momentum \cite{Naumov2013}.

The content of this work has been organized as follows: In section
two, considering neutrinos as Majorana fermions, we show how is
possible to obtain the standard two-flavor neutrino oscillation
probabilities in the context of a quantum field theory treatment,
for which the mass eigenstate are described by plane waves; in
section three, we extend the plane wave treatment presented in the
previous section for the case in which the mass eigenstates are
described by localized wave packets; in section four, we study the
effects of spreading of wave packets by performing an expansion of
the energy until second order in the momentum and we obtain the
standard time-integrated neutrino oscillation probabilities; in
section five, we study the effects originated in dispersion with
time on spreading of wave packets for the time-integrated neutrino
oscillation probabilities by performing an expansion of the energy
until third order in the momentum; finally in section six we
present some conclusions.

\section{Neutrino oscillation probabilities using plane waves}

The standard two-flavor neutrino oscillation probabilities were
obtained in the context of a treatment developed in the canonical
formalism of quantum field theory for the case in which neutrinos
were described as Majorana fermions and neutrino mass eigenstates
were described by plane waves \cite{PQ1}. In this treatment, the
flavor neutrinos were considered as superpositions of mass
eigenstates with specific momenta. For the case of the
relativistic limit $(L \simeq T)$ and after including a
normalization constant, the standard plane wave expressions for
the neutrino oscillation probabilities \cite{PQ1} are obtained
\begin{align}\label{eq:nudpt1}
  P_{\nu_{e}}^{PW}(L)=1-\sin^2\left[2\theta_{12}\right]
  \sin^2\left[\frac{\Delta m_{12}^2}{4E}L\right],
\end{align}
\begin{align}\label{eq:edpt1}
  P_{\nu_{\mu}}^{PW}(L)=\sin^2\left[2\theta_{12}\right]
  \sin^2\left[\frac{\Delta m_{12}^2}{4E}L\right],
\end{align}
where $\Delta m_{12}^2\equiv m_2^2-m_1^2$, $E$ is the energy of
the neutrino, $L$ is the distance between the neutrino source and
the detector and $\sin^2[2\theta_{12}]$ is given by
\begin{equation}\label{eq:thlamb}
\sin^2[2\theta_{12}]=\frac{4\Lambda_L^2}{(1+\Lambda_L^2)^2},
\end{equation}
with $\theta_{12}$ representing the mixing angle between the two
mass eigenstates in the vacuum. In the last expression $\Lambda_L$
is
\begin{equation} \label{eq:Lambda}
 \Lambda_L
 =\frac{m_{\nu_{\mu_L}}-m_{\nu_{e_L}}+R_L}{2m_{\nu_{e_L}\nu_{\mu_L}}},
\end{equation}
where $R_L$ is defined by means of
\begin{equation} \label{eq:Lambda}
 R_L^2=(m_{\nu_{e_L}}-m_{\nu_{\mu_L}})^2+4m_{\nu_{e_L}\nu_{\mu_L}}^2,
\end{equation}
and the parameters $m_{\nu_{e_L}}$, $m_{\nu_{\mu_L}}$ and
$m_{\nu_{e_L}\nu_{\mu_L}}$ are related with the masses $m_1$ and
$m_2$ of the neutrino fields $\nu_1$ and $\nu_2$ (with definite
masses) through the following relations
\begin{align}
 m_{1}&=\frac{1}{2}(m_{\nu_{e_L}}+m_{\nu_{\mu_L}}-R_L),\\
 m_{2}&=\frac{1}{2}(m_{\nu_{e_L}}+m_{\nu_{\mu_L}}+R_L)e^{-i\alpha_L},
\end{align}
being $\alpha_L$ the Majorana complex phase \cite{PQ1}.

The expressions (\ref{eq:nudpt1}) and (\ref{eq:edpt1}) are
obtained starting from the oscillation probabilities defined by
\begin{equation} \label{eq:nudpt2}
P_{\nu_{e}}^{PW}(L)=
\left|\bra{0}\des{\nu}{e_L}(x)\ket{\nu_{e_L}^{PW}(x_0)}\right|^2,
\end{equation}
\begin{equation} \label{eq:edpt2}
P_{\nu_{\mu}}^{PW}(L)=
\left|\bra{0}\hat\nu_{\mu_L}(x)\ket{\nu_{e_L}^{PW}(x_0)}\right|^2,
\end{equation}
in such a way that in the space-time production point ($x_0$) the
initial left-handed neutrino flavor state (electron neutrino
flavor) $\ket{\nu_{e_L}^{PW}(x_0)}$ is defined by the following
superposition of the mass eigenstates $\ket{\nu_{1}^{PW}(x_0)}$
and $\ket{\nu_{2}^{PW}(x_0)}$
\begin{align}\label{eq:EI}
\ket{\nu_{e_L}^{PW}(x_0)}=\sum_{h=\pm 1}\frac{\Lambda_L
}{\sqrt{1+\Lambda_L^2}}\ket{\nu_{1}^{PW}(x_0)}+\frac{e^{-i\alpha_L}}{\sqrt{1+
\Lambda_L^2}}\ket{\nu_{2}^{PW}(x_0)},
\end{align}
where a sum over helicities is taken in the superposition. The
mass eigenstates $\ket{\nu_{1}^{PW}(x_0)}$ and
$\ket{\nu_{2}^{PW}(x_0)}$ involved in (\ref{eq:EI}) are obtained
using plane waves from the vacuum state $\ket{0}$ as
\begin{equation} \label{eq:nms}
\ket{\nu_{a}^{PW}(x_0)}=Ae^{ip_ax_0}\crea{a}{a}(\vec{p}_a,h)\ket{0}
\end{equation}
where $A$ is a normalization constant, $\crea{a}{a}$ is the
creation operator of a neutrino of defined mass, $x_0$ is the
space-time point where this neutrino is created,
$p_a=(E_a,\vec{p}_a)$ is the four-momentum of the mass eigenstates
and $a=1,2$. We have assumed that each mass eigenstate involved in
(\ref{eq:EI}) has associate a specific four-momentum.

In the oscillation probabilities (\ref{eq:nudpt2}) and
(\ref{eq:edpt2}), the flavor neutrino field operator $\hat
\nu_\alpha$ is defined as a superposition of field operators of
neutrinos with defined mass $\hat \nu_a$ by means of the
expression
\begin{align}\label{eq:rot}
 \des{\nu}{\alpha}(x)=\sum_aU_{L_{\alpha a}}\des{\nu}{a}(x),
\end{align}
where $\alpha=e_L,\mu_L$ and $U_L$ is an unitarian rotation matrix
given by \cite{PQ1}
\begin{align}\label{eq:MUD}
 U_L=\frac{1}{\sqrt{1+\Lambda_L^2}}\begin{pmatrix}
                                \Lambda_L & e^{-i\alpha_L}\\
                -1 & \Lambda_L e^{-i\alpha_L}
                               \end{pmatrix}.
\end{align}
The field operators of neutrinos with defined mass $\hat \nu_a$
involved in (\ref{eq:rot}) are defined as \cite{PQ1}
\begin{eqnarray}\label{eq:FEOFDM}
\hat\nu_a(x)=\int \frac{d^3p}{(2\pi)^{3/2}(2E_a)^{1/2}}\sum_{
h=\pm1}&\left[\sqrt{E_a-h
\abs{\vec{p}}}\,\des{a}{a}(\vec{p},h)\chi^{h}(\vec{p})e^{-ip\cdot
x}\right.\nonumber
\\ &\left.-h\sqrt{E_a+h\abs{\vec{p}}}\,\crea{a}{a}(\vec{p},h)
\chi^{-h}(\vec{p})e^{ip\cdot x}\right],
\end{eqnarray}
where $E_a^2=\abs{\vec{p}}^2+m_a^2$ is the energy of the neutrino
field with defined mass and $\chi^h(\vec{p})$ are Majorana spinors
with helicity eigenvalues $\pm 1$.

\section{Neutrino oscillations using the wave packets formalism}

In this section, we will extend the plane wave treatment for
neutrino oscillations that we have have presented briefly in
section two, for the case in which neutrino mass eigenstates are
described by localized wave packets. In this treatment, where
neutrinos are considered as Majorana fermions, we do not focus on
the study of the details of the interaction processes in which
neutrinos are produced and detected. Here, on the other hand, it
is assumed that wave packets describing mass eigenstates are
localized and the coefficients of their superpositions are given
by the elements of the unitarian rotation matrix $U_L$ given by
(\ref{eq:MUD}). The matrix $U_L$ establishes a relationship
between the flavor and mass eigenstates bases of neutrino fields
in vacuum.

There are different reasons for understand why the description of
mass eigenstates using wave packets is most appropriate to study
the neutrino oscillations with respect to the description from
plane waves \cite{CK93,RM02,GK07}. Some of these reasons are that
the neutrino source and the detector are localized and there
exists a spread for the neutrino momentum \cite{CK93}.
Additionally, we have to keep in mind that plane waves localized
in some point $x_0$ are in an obvious contradiction with the
Heisenberg uncertainty principle. Here, we take into account these
reasons when we describe the neutrino mass eigenstates in terms of
superpositions of localized wave packets. To do the last, we first
consider that in a point $x_0\equiv x_0^\mu=(t_0,\vec{r}_0)$ is
created a left-handed electron neutrino which is described by the
following superposition of neutrino mass eigenstates
$\ket{\nu_{1}^{WP}(x_0)}$ and $\ket{\nu_{2}^{WP}(x_0)}$
\begin{align}\label{eq:EIWP}
\ket{\nu_{e_L}^{WP}(x_0)}=\sum_{h=\pm 1}\frac{\Lambda_L
}{\sqrt{1+\Lambda_L^2}}\ket{\nu_{1}^{WP}(x_0)}+\frac{e^{-i\alpha_L}}{\sqrt{1+
\Lambda_L^2}}\ket{\nu_{2}^{WP}(x_0)}.
\end{align}

In contrast to the expression (\ref{eq:nms}), now the mass
eigenstates involved in (\ref{eq:EIWP}) are written in term of
localized wave packets in the form
\begin{align}\label{eq:nmswp}
  \ket{\nu_{a}^{WP}(x_0)} = A\,\int \frac{d^3p}{(2\pi)^{3/2}}e^{-i(E_at_0-
  \vec{p}\cdot\vec{r}_0)}\psi_a(\vec{p},\langle\vec{p}_a\rangle)
  \,\crea{a}{a}(\vec{p},h)\ket{0},
\end{align}
where $A$ is a normalization constant,
$\psi_a(\vec{p},\langle\vec{p}_a\rangle)$ is a probability density
function which depends on the momentum $\vec{p}_a$ and the average
momentum $\langle\vec{p}_a\rangle$, and $a=1,2$. In general,
$\psi_a(\vec{p},\langle\vec{p}_a\rangle)$ may take any form, but
it is usually approximated by a Gaussian distribution assuming
that is peaked around the average momentum
\cite{Giunti1991,Giunti1993,Giunti1998-1,Giunti1998-2,Beuthe2002,Giunti2002,Giunti2003,Beuthe2003,
Giunti2004,Giunti2007,Bilenky2011,Bilenky2011-2,Bernardini2004,Bernardini2005}.
In this work, we define the probability density function as
\cite{GK07}
\begin{align}\label{eq:DM}
\psi_a(\vec{p},\langle\vec{p}_a\rangle)\approx
(2\pi)^{-3/4}[\text{Det}\,\Gamma]^{1/4}\exp\left[-\frac{1}{4}(\vec{p}-
\langle\vec{p}_a\rangle)^k\Gamma_{kj}(\vec{p}-\langle\vec{p}_a\rangle)^j
\right],
\end{align}
in such a way that
$\psi_a=\psi_a(\vec{p},\langle\vec{p}_a\rangle)$ satisfies the
conditions \cite{GK07}
\begin{align}
\left.\parc{\ln \psi_a}{p}\right|_{\vec{p}=\langle\vec{p}_a\rangle}&=0,\\
\left.\frac{\partial^2\ln \psi_a}{\partial p^k\partial
p^j}\right|_{\vec{p}
=\langle\vec{p}_a\rangle}&=-\frac{1}{2}\Gamma_{kj}
\end{align}
In the probability density function given by (\ref{eq:DM}), we
have taken the convention of summation over the Latin repeated
index $k$ and $j$. Additionally, we have assumed that the
dispersion over the mass eigenstates $\ket{\nu_{1}^{WP}(x_0)}$ and
$\ket{\nu_{1}^{WP}(x_0)}$ is the same, because these eigenstates
are created simultaneously by the same weak production process.
This fact is the reason that justifies why the matrix
$\Gamma_{kj}$ is identic for both mass eigenstates. Moreover, it
is important to note that this matrix is symmetric and the
eigenvalues of its inverse $\Gamma_{kj}^{-1}$
are the squares of the widths in the momentum space \cite{GK07}.\\

The oscillation probabilities between two flavor neutrinos using
wave packets are
\begin{equation} \label{eq:nudpt3}
P_{\nu_{e}}^{WP}((T,L)=
\left|\bra{0}\des{\nu}{e_L}(x)\ket{\nu_{e_L}^{WP}(x_0)}\right|^2,
\end{equation}
\begin{equation} \label{eq:edpt3}
P_{\nu_{\mu}}^{WP}(T,L)=
\left|\bra{0}\hat\nu_{\mu_L}(x)\ket{\nu_{e_L}^{WP}(x_0)}\right|^2,
\end{equation}
where $\ket{\nu_{e_L}^{WP}(x_0)}$ given by (\ref{eq:EIWP}) is a
superposition of mass eigenstate wave packets
$\ket{\nu_{1}^{WP}(x_0)}$ and $\ket{\nu_{1}^{WP}(x_0)}$) in the
creation point. The expressions (\ref{eq:nudpt3}) and
(\ref{eq:edpt3}) describe, respectively, the probabilities of
finding an electron neutrino $(\nu_e)$ and a muon neutrino
$(\nu_\mu)$ at a distance $L$ in a time $T$ of the creation point
$x_0$. In the calculation of the transition probabilities
(\ref{eq:nudpt3}) and (\ref{eq:edpt3}), the relativistic
dispersion relation is approximated by means of an expansion
around the average momentum of the wave packets
$\langle\vec{p}_a\rangle$ \cite{Giunti2002}
\begin{align}\label{eq:STE}
E_a(\vec{p})\approx \bar E_a+\vec{v}_a\cdot(\vec{p}-\langle
\vec{p}_a\rangle)+
\frac{1}{2}(\vec{p}-\langle\vec{p}_a\rangle)^k\Omega_{kj}
(\vec{p}-\langle\vec{p}_a\rangle)^j+\cdots,
\end{align}
where \cite{Giunti2002}
\begin{align}
\bar E_a&\equiv E(\langle\vec{p}_a\rangle)=\sqrt{\langle\vec{p}_a
\rangle^2+m_a^2},\label{eq:Ener}\\
v_a^k&=\left.\parc{E(\vec{p}_a)}{p_k}\right|_{\vec{p}=\langle\vec{p}_a
\rangle}=\frac{\langle\vec{p}_a\rangle^k}{\bar E_a},\label{eq:Velo}\\
\Omega_{kj}^a&=\left.\frac{\partial^2E(\vec{p}_a)}{\partial
p^k\partial
p^j}\right|_{\vec{p}=\langle\vec{p}_a\rangle}=\frac{1}{\bar
E_a}\corc{\delta_{kj}-v_k^av_j^a}.
\end{align}
Additionally, it is possible to write that \cite{GK07}
\begin{align}
\sqrt{\frac{E_a(\vec{p})\pm h
\abs{\vec{p}}}{2E_a(\vec{p})}}&\approx
\sqrt{\frac{\bar E_a\pm h \abs{\langle\vec{p}_a\rangle}}{2\bar E_a}},\\
\chi^h(\vec{p})&\approx\chi^h(\langle\vec{p}_a\rangle).
\end{align}
The highest power of the momentum $(\vec p
-\langle\vec{p}_a\rangle)$ in the expansion of the energy given by
(\ref{eq:STE}) determines the two different cases which we will
studied below: (i) If the highest power is taken until second
order, the effects of the spreading of the mass eigenstates wave
packets lead to the standard time-integrated neutrino oscillation
probabilities obtained using the wave packet formalism
\cite{Giunti1993,Giunti1998-1,Giunti1998-2,Beuthe2002,Giunti2002,Giunti2003,Beuthe2003,
Giunti2004,Giunti2007,Bilenky2011,Bilenky2011-2}; (ii) If the
highest power is taken until third order, the effects originated
in dispersion with time on the spreading of wave packets for the
two-flavor neutrino oscillation probabilities are observed as a
factor that suppress the standard time-integrated neutrino
oscillation probabilities.

\section{Expansion of the energy until second order in the momentum}

In this section, we will obtain the standard time-integrated
neutrino oscillation probabilities using the wave packet
formalism. To do it, we expand the energy given by (\ref{eq:STE})
up to second order in the power series of $(\vec p
-\langle\vec{p}_a\rangle)$ \cite{Giunti2002}. This fact is
justified taking into account that the width of the wave packets
is very narrow around the average momentum. For this case, the
matrix $\Gamma_{kj}$ can be diagonalized by means of an orthogonal
transformation, {\it i. e.} a rotation \cite{Giunti2002,GK07}.
Given that the expansion of the energy does not change this
rotation, without lost of generality we can take a reference frame
where the matrix is diagonal
\begin{align}\label{eq:MGD}
\Gamma_{kj}=\frac{1}{\sigma_p^2}\delta_{kj},
\end{align}
with $\sigma_p$ representing the width of the wave packets in the
momentum space. We assume that the width has the same value for
each of the dimensions of the momentum space, due to the wave
packets are taken as isotropic. Additionally, we define the width
of the wave packets in the coordinate space $\sigma_r$ through the
uncertainty relation
\begin{align}
\sigma_r\sigma_p=\frac{1}{2}.
\end{align}

If the energy given by (\ref{eq:STE}) is substituted in
(\ref{eq:nudpt3}) and (\ref{eq:edpt3}), keeping up to the second
order in the power series of $(\vec p -\langle\vec{p}_a\rangle)$,
we obtain that the neutrino oscillation probabilities are written
as
\begin{eqnarray}
P_{\nu_e}^{SWP}(T,L)=&\frac{1}{(2\pi\sigma_r^2)^{1/2}}\frac{1}{(1+\Lambda^2)^2}
\left\{ \Lambda^4\,\exp[-\lambda_1\phi_{1}^{S}(T)] \right.
\nonumber \\&\left. + \exp[-\lambda_2\phi_{2}^{S}(T)]+
\Lambda^2\aleph\,\exp[-\lambda_3\phi_{3}^{S}(T)] \right\}
\label{eq:nudpt4a},
\end{eqnarray}
\begin{eqnarray}
P_{\nu_\mu}^{SWP}(T,L)=&\frac{1}{(2\pi\sigma_r^2)^{1/2}}\frac{1}{(1+\Lambda^2)^2}
\left\{ \Lambda^2\,\exp[-\lambda_1\phi_{1}^{S}(T)] \right.
\nonumber \\&\left. + \Lambda^2\,\exp[-\lambda_2\phi_{2}^{S}(T)]-
\Lambda^2\aleph\,\exp[-\lambda_3\phi_{3}^{S}(T)] \right\}
\label{eq:edpt4a},
\end{eqnarray}
where $\lambda_1=\lambda_2 = 1/2\sigma_r^2$, $\lambda_3
=1/4\sigma_r^2$, $T=t-t_0$, $L=\abs{\vec{r}-\vec{r}_0}$ and the
functions in the arguments of the exponentials are given by
\begin{align}
\phi_{1}^{S}(T)=&(L-v_1T)^2,\label{eq:defxS1}\\
\phi_{2}^{S}(T)=&(L-v_2T)^2,\label{eq:defxS2}
\end{align}
\begin{equation}
\phi_{3}^{S}(T)=(L-v_1T)^2+(L-v_2T)^2-i4\sigma_r^2(\bar E_1-\bar
E_2)T+i4\sigma_r^2(\bar p_1-\bar p_2)L,
\end{equation}
with $v_a=\abs{\vec{v}_a}$. The quantity $\aleph$ that appears in
the oscillation probabilities (\ref{eq:nudpt4a}) and
(\ref{eq:edpt4a}) is written as
\begin{align}\label{eq:defx}
\aleph=\frac{1}{(\bar E_1\bar E_2)^{1/2}}\sum_h\sqrt{(\bar
E_1-h\abs{\langle\vec{p}_1\rangle})(\bar
E_2-h\abs{\langle\vec{p}_2\rangle})}.
\end{align}
Now, we take into account the fact that in the atmospheric and
reactor neutrino oscillation experiments it is only possible to
measure the distance between the neutrino source and the detector
$L$, while the neutrino propagation time $T$ is unknown
\cite{Giunti1991,Beuthe2003,Bilenky2011-2}. However, in the case
of the accelerator neutrino experiments (for instance K2K, MINOS,
OPERA) it is possible to measure the neutrino propagation time $T$
\cite{Bilenky2011-2}. By this reason, if we focus only on the case
of atmospheric and reactor neutrino oscillation experiments, then
it is necessary the elimination of the time dependence presents in
(\ref{eq:nudpt4a}) and (\ref{eq:edpt4a}). This last can be
performed, if we take the average on the time of the expressions
(\ref{eq:nudpt4a}) and (\ref{eq:edpt4b}) in the following form
\begin{align}
P_{\nu_e}^{SWP}(L)&=\int P_{\nu_e}^{SWP}(T,L) dT,\label{eq:nudpt4b}\\
P_{\nu_\mu}^{SWP}(L)&=\int P_{\nu_\mu}^{SWP}(T,L)
dT,\label{eq:edpt4b}
\end{align}
Time integrations can be performed using both Gaussian integration
and the Laplace approximation method. After the time integrations
are performed, we obtain from (\ref{eq:nudpt4b}) and
(\ref{eq:edpt4b}) the following time-integrated neutrino
oscillation probabilities
\begin{equation} \label{eq:nudpt4c}
P_{\nu_e}^{SWP}(L)=\frac{1}{(1+\Lambda^2)^2}\left\{\frac{
\Lambda^4}{v_1}+\frac{1}{v_2}+\Lambda^2\Xi\left(
\frac{2}{v_1^2+v_2^2} \right)^{1/2} \exp\left[i
f_{1}^{S}-f_{2}^{S}\right]\right\},
\end{equation}
\begin{equation} \label{eq:edpt4c}
P_{\nu_\mu}^{SWP}(L)=\frac{1}{(1+\Lambda^2)^2}\left\{\frac{
\Lambda^2}{v_1}+\frac{\Lambda^2}{v_2}-\Lambda^2\Xi\left(
\frac{2}{v_1^2+v_2^2} \right)^{1/2} \exp\left[i
f_{1}^{S}-f_{2}^{S}\right]\right\},
\end{equation}
where
\begin{align}
f_{1}^{S}&=(\bar E_1 - \bar
E_2)\frac{v_1+v_2}{v_1^2+v_2^2}L-(\bar p_1 -\bar p_2)L,\\
f_{2}^{S}&=\frac{(v_1 - v_2)^2}{v_1^2 +
v_2^2}\frac{L^2}{4\sigma_r^2} + \frac{(\bar E_1- \bar
E_2)^2}{v_1^2 + v_2^2} \sigma_r^2.
\end{align}
We have explicitly proved that if the average on the time in the
expressions (\ref{eq:nudpt4b}) and (\ref{eq:edpt4b}) is performed
using Gaussian integration, the results are the same as those
obtained using the Laplace approximation method. In both cases, we
have obtained the oscillation probabilities given by
(\ref{eq:nudpt4c}) and (\ref{eq:edpt4c}). The functional form of
the oscillation probabilities (\ref{eq:nudpt4c}) and
(\ref{eq:edpt4c}) is in agreement with the one previously obtained
by Giunti, Kim and Lee. These authors used a quantum mechanics
treatment in which flavor neutrinos were described by a
superposition of mass eigenstates wave packets \cite{Giunti1991}.

In order to obtain from (\ref{eq:nudpt4c}) and (\ref{eq:edpt4c})
expressions for the oscillation probabilities in the relativistic
limit, the following relativistic approximations are used
\cite{Giunti1998-1,Giunti1998-2,Giunti2003}
\begin{align}
{\bar E_a} \simeq {\bar E} + \xi \frac{m_a^2}{2{\bar E}},\label{eq:relap1}\\
{\bar p_a} \simeq {\bar E} + (1- \xi) \frac{m_a^2}{2{\bar E}},\label{eq:relap2}\\
{v_a} \simeq 1 - \frac{m_a^2}{2{\bar E}},\label{eq:relap3}
\end{align}
where $\xi$ is a dimensionless coefficient, typically of order
unity, that depends of the neutrino production process and $\bar
E$ is the neutrino energy determined by the kinematics of the
production process for a massless neutrino. After the relativistic
limit is taken, we obtain from (\ref{eq:nudpt4c}) and
(\ref{eq:edpt4c}) the standard time-integrated neutrino
oscillation probabilities
\begin{equation} \label{eq:nudpt4d}
P_{\nu_e}^{SWP}(L)=1-\frac{1}{2}\sin^2[2\theta_{12}]\left\{1-
\exp\left[i\,2\pi\frac{L}{L_{osc}}-\left(\frac{L}{L_{coh}}\right)^2-
2\pi^2
\xi^2\left(\frac{\sigma_r}{L_{osc}}\right)^2\right]\right\},
\end{equation}
\begin{equation} \label{eq:edpt4d}
P_{\nu_\mu}^{SWP}(L)=\frac{1}{2}\sin^2[2\theta_{12}]\left\{1-
\exp\left[i\,2\pi\frac{L}{L_{osc}}-\left(\frac{L}{L_{coh}}\right)^2-
2\pi^2
\xi^2\left(\frac{\sigma_r}{L_{osc}}\right)^2\right]\right\},
\end{equation}
where $L_{osc}$ is the oscillation length and $L_{coh}$ is the
coherence length given by
\begin{align}
L_{osc}=&\frac{4\pi\bar E}{\Delta m_{12}^2},\label{eq:LOsc}\\
L_{coh}=&\frac{4\sqrt{2}\bar E^2}{\Delta
m_{12}^2}\sigma_r,\label{eq:LCoh}
\end{align}
in agreement with the corresponding lengths very well known in the
literature
\cite{Giunti1991,Giunti1993,Giunti1998-1,Giunti1998-2,Beuthe2002,Giunti2002,Giunti2003,Beuthe2003,
Giunti2004,Giunti2007,Bilenky2011,Bilenky2011-2}. To write the
expressions (\ref{eq:nudpt4d}) and (\ref{eq:edpt4d}), we have used
the definition of $\sin^2[2\theta_{12}]$ in terms of the parameter
$\Lambda$ given by (\ref{eq:thlamb}), where $\theta_{12}$ is the
mixing angle in the vacuum between the two mass eigenstates.
Specifically, the probability (\ref{eq:nudpt4d}) represents the
survival probability that an electron neutrino $(\nu_e)$ be
detected at a distance $L$ in a time $T$ of the creation point
$x_0=(0,\vec r_0)$, where by simplicity $t_0=0$. On the other
hand, the probability (\ref{eq:edpt4d}) represents the probability
of oscillation from an electron neutrino $(\nu_e)$ created by the
source at point $x_0$ to a muon neutrino $(\nu_\mu)$ measured by
the detector at a distance $L$ in time $T$. The dependence of the
oscillation probability (\ref{eq:edpt4d}) respect to the mixing
angle $\theta_L$ is in agreement with the reported by Bernardini
and De Leo. These authors studied the effects of positive and
negative energy components of mass eigenstate wave packets on the
two-flavor neutrino oscillation probabilities
\cite{Bernardini2004,Bernardini2005}.

The dependence of the time-integrated oscillation probabilities
(\ref{eq:nudpt4d}) and (\ref{eq:edpt4d}) respect to $L_{osc}$ and
$L_{coh}$ is in agreement with the reported in the literature
\cite{Giunti1991,Giunti1993,Giunti1998-1,Giunti1998-2,Beuthe2002,Giunti2002,Giunti2003,Beuthe2003,
Giunti2004,Giunti2007,Bilenky2011,Bilenky2011-2}. The first term
in the argument of the exponentials in (\ref{eq:nudpt4d}) and
(\ref{eq:edpt4d}) is the standard oscillation phase proportional
to the propagation distance $L$. The second term in the argument
of the exponentials is called the damping factor
\cite{Giunti1991}. This term implies a quadratical decrease of the
oscillation probabilities with the distance $L$ and determines how
far the oscillations take place \cite{Giunti1998-1}. For $L \gg
L_{coh}$, the interference of the neutrino mass eigenstates is
suppressed and the oscillations due to $L_{osc}$ disappear
\cite{Giunti1998-1}. This behavior can be partially originated in
the progressive separation of mass eigenstates wave packets
propagating in space \cite{Beuthe2003}. The third term in the
argument of the exponential is called the localization factor
\cite{Giunti1998-1} and this term implies that
\begin{align}\label{eq:cond1}
\sigma_r < \frac{L_{osc}}{\sqrt{2}\pi}=\frac{\sqrt{8}{\bar
E}}{\Delta m_{12}^2},
\end{align}
meaning that the neutrino production process, which is
characterized by the width of the wave packets $\sigma_r$, is
localized in a region much smaller that the oscillation length
$L_{osc}$ \cite{Giunti1998-1}. Thus, this factor does not depends
on the distance $L$.

As can be observed from (\ref{eq:LOsc}) and (\ref{eq:LCoh}), the
oscillation length and the coherence length are related by
\begin{equation}
L_{coh}=\frac{\sqrt{2}}{\pi}\sigma_r \bar E \, L_{osc},
\end{equation}
showing that the coherence length is much larger than the
oscillation length \cite{Bilenky2011-2}. The maximum number of
oscillations can be obtained from these lengths as
\cite{Giunti1991}
\begin{equation}\label{eq:Numos}
N_{osc}=\frac{L_{coh}}{L_{osc}}= \frac{\sqrt{2}}{\pi} \sigma_r
\bar E.
\end{equation}

Because in the neutrino oscillation experiments one has $L \simeq
L_{osc}$, then the term $\exp\left[
-\left(\frac{L}{L_{coh}}\right)^2 \right]$ is nearly equal to one
\cite{Bilenky2011-2}. Additionally, it is easy to show that $\mid
v_1 - v_2 \mid L_{coh} \simeq \frac{\mid \Delta m_{12}^2 \mid}{2
\bar E} \sim \sigma_r$, then $L_{osc} \gg \sigma_r$ and thus the
term $\exp\left[ -2 \pi^2 \xi^2
\left(\frac{\sigma_r}{L_{osc}}\right)^2 \right]$ is also nearly
equal to one \cite{Bilenky2011-2}. In this form, the
time-integrated oscillation probabilities (\ref{eq:nudpt4d}) and
(\ref{eq:edpt4d}) can be written as \cite{Bilenky2011-2}
\begin{equation} \label{eq:nudpt4e}
P_{\nu_e}^{SWP}(L)=1-\frac{1}{2}\sin^2[2\theta_{12}]\left\{1-
\cos\left[2\pi\frac{L}{L_{osc}}\right]\right\},
\end{equation}
\begin{equation} \label{eq:edpt4e}
P_{\nu_\mu}^{SWP}(L)=\frac{1}{2}\sin^2[2\theta_{12}]\left\{1-
\cos\left[2\pi\frac{L}{L_{osc}}\right]\right\},
\end{equation}
that reduce to the standard neutrino oscillation probabilities
(\ref{eq:nudpt1}) and (\ref{eq:edpt1}) obtained using the plane
wave formalism.

\section{Expansion of the energy until third order in the momentum}

In this section, we will study the effects originated in
dispersion with time on spreading of wave packets for the
time-integrated two-flavor neutrino oscillation probabilities by
expanding the energy given by (\ref{eq:STE}) up to third order in
the momentum $(\vec p -\langle\vec{p}_a\rangle)$. For this case,
we take a reference frame where the matrix $\Gamma_{kj}$ is
diagonal and identical to (\ref{eq:MGD}). Substituting the energy
given by (\ref{eq:STE}) in (\ref{eq:nudpt3}) and (\ref{eq:edpt3}),
keeping up to the third order in the power series of $(\vec p
-\langle\vec{p}_a\rangle)$, we obtain that the neutrino
oscillation probabilities are written as
\begin{eqnarray}
P_{\nu_e}^{DWP}(T,L)=&\frac{1}{(2\pi\sigma_r^2)^{1/2}}\frac{1}{(1+\Lambda^2)^2}
\left\{\frac{\Lambda^4}{g_{1}^{D}(T)}\exp[-\lambda_1\phi_{1}^{D}(T)]
\right. \nonumber\\ &\left. +
\frac{1}{g_{2}^{D}(T)}\exp[-\lambda_2\phi_{2}^{D}(T)]+
\frac{\Lambda^2\aleph}{g_{3}^{D}(T)}\exp[-\lambda_3\phi_{3}^{D}(T)]
\right\} \label{eq:nudpt5a},
\end{eqnarray}
\begin{eqnarray}
P_{\nu_\mu}^{DWP}(T,L)=&\frac{1}{(2\pi\sigma_r^2)^{1/2}}\frac{1}{(1+\Lambda^2)^2}
\left\{\frac{\Lambda^2}{g_{1}^{D}(T)}\exp[-\lambda_1\phi_{1}^{D}(T)]+
\right. \nonumber\\ &\left.
\frac{\Lambda^2}{g_{2}^{D}(T)}\exp[-\lambda_2\phi_{2}^{D}(T)]-
\frac{\Lambda^2\aleph}{g_{3}^{D}(T)}\exp[-\lambda_3\phi_{3}^{D}(T)]
\right\} \label{eq:edpt5a},
\end{eqnarray}
with $\aleph$ given by (\ref{eq:defx}), $\lambda_1=\lambda_2 =
1/2\sigma_r^2$, $\lambda_3 =1/4\sigma_r^2$, and the functions in
the arguments of the exponentials are
\begin{align}
\phi_{1}^{D}(T)=&\frac{(L-v_1T)^2}{1+\frac{T^2}{(T_{1}^{L})^2}}, \label{eq:phi1}\\
\phi_{2}^{D}(T)=&\frac{(L-v_2T)^2}{1+\frac{T^2}{(T_{2}^{L})^2}},
\label{eq:phi2}
\end{align}
\begin{equation}
\phi_{3}^{D}(T)=\frac{(L-v_1T)^2}{1-\frac{iT}{T_{1}^{L}}}+
\frac{(L-v_2T)^2}{1-\frac{iT}{T_{2}^{L}}}-i4\sigma_r^2(\bar
E_1-\bar E_2)T+i4\sigma_r^2(\bar p_1-\bar p_2)L, \label{eq:phi3}
\end{equation}
while the functions $g_{1}^{T}(T)$, $g_{2}^{T}(T)$ and
$g_{3}^{T}(T)$ are
\begin{align}
g_{1}^{D}(T)=&\left(1+\frac{T^2}{(T^{T})^2}\right)
\left(1+\frac{T^2}{(T_{1}^{L})^2}\right)^{1/2}, \label{eq:fg1}\\
g_{2}^{D}(T)=&\left(1+\frac{T^2}{(T^{T})^2}\right)
\left(1+\frac{T^2}{(T_{2}^{L})^2}\right)^{1/2},\label{eq:fg2}
\end{align}
\begin{equation}
g_{3}^{D}(T)=\left(1-i\frac{T}{T^{T}}\right)
\left(1+i\frac{T}{T^{T}}\right)\left(1-i\frac{T}{T_{1}^{L}}\right)^{1/2}
\left(1+i\frac{T}{T_{2}^{L}}\right)^{1/2}, \label{eq:fg3}
\end{equation}
where we have defined the longitudinal dispersion time $T_{a}^{L}$
as $T_{a}^{L}=2\bar E^3\sigma_r^2/m_a^2$, with $a=1,2$, while the
transversal dispersion time $T^{T}$ has been defined as
$T^{T}=2\bar E\sigma_r^2$. The longitudinal dispersion time in
neutrino oscillations was initially defined in the context of a
quantum mechanics treatment \cite{Giunti1991}. This time was
posteriorly considered in the context of a quantum field theory
treatment of neutrino oscillations \cite{Beuthe2002,Beuthe2003}.
Most recently, the transversal and longitudinal times that we have
defined here were considered in the context of a theory of wave
packets in which the energy that appears in the wave packets is
expanded until third order in the momentum \cite{Naumov2013}.

Because the longitudinal and transversal times are related as
$T^L_a=\frac{\bar E^2}{m_a^2} T^L$, we observe that $T^L_a \gg
T^T$ and these two very separated times can be used to define
three dispersion regimes: {\it (i)} The minimum dispersion regime
is defined for times $T$ that satisfy $T<T^T$;  {\it (ii)} the
transversal dispersion regime for $T^T<T<T^L$; {\it (iii)} the
longitudinal dispersion regime for $T>T^L$. These three dispersion
regimes were equivalently considered previously by using the
distance between the neutrino source and the detector $L$ as the
quantity to define these regimes \cite{Beuthe2003}.

We observe in (\ref{eq:nudpt5a}) and (\ref{eq:edpt5a}) the
existence of two different longitudinal dispersion times
$T_{1}^{L}$ and $T_{2}^{L}$. For simplicity, we will work in the
limit in which the masses are nearly degenerate $m_1=m_2=\bar m$.
For this limit, it is possible to consider $T_{1}^{L}= T_{2}^{L}$
and to work with only one longitudinal dispersion time defined by
$T^{L}=2\bar E^3\sigma_r^2/\bar m^2$, with $\bar m$ the mass in
the degenerate limit \cite{Beuthe2003}. For the nearly degenerate
limit, the functions given by the expressions (\ref{eq:fg1}),
(\ref{eq:fg2}) and (\ref{eq:fg3}) are written as
\begin{align}
g_{1}^{D}(T)=g_{2}^{D}(T)=g_{3}^{D}(T)=&\left(1+\frac{T^2}{(T^{T})^2}\right)
\left(1+\frac{T^2}{(T^{L})^2}\right)^{1/2}, \label{eq:fg4}
\end{align}

In the next, we will study for the three dispersion regimes
previously defined the effects originated in dispersion with time
on the spreading of wave packets for the two-flavor neutrino
oscillation probabilities.

\subsection{Spreading in the minimum dispersion regime}

In the minimum dispersion regime $T<T^T$ and for the limit of
nearly degenerate masses, the oscillation probabilities
(\ref{eq:nudpt5a}) and (\ref{eq:edpt5a}) can be written as
\begin{eqnarray}
P_{\nu_e}^{MDWP}(T,L)=&\frac{1}{(2\pi\sigma_r^2)^{1/2}}\frac{1}{(1+\Lambda^2)^2}
\left\{ \Lambda^4\,\exp[-\lambda_1\phi_{1}^{S}(T)]\right.
\nonumber \\ &\left. + \exp[-\lambda_2\phi_{2}^{S}(T)]+
\Lambda^2\aleph\,\exp[-\lambda_3\phi_{3}^{M}(T)] \right\}
\label{eq:nudpt5b},
\end{eqnarray}
\begin{eqnarray}
P_{\nu_\mu}^{MDWP}(T,L)=&\frac{1}{(2\pi\sigma_r^2)^{1/2}}\frac{1}{(1+\Lambda^2)^2}
\left\{ \Lambda^2\,\exp[-\lambda_1\phi_{1}^{S}(T)]\right.
\nonumber \\ &\left. + \Lambda^2\,\exp[-\lambda_2\phi_{2}^{S}(T)]-
\Lambda^2\aleph\,\exp[-\lambda_3\phi_{3}^{M}(T)] \right\}
\label{eq:edpt5b},
\end{eqnarray}
with the functions $\phi_{1}^{S}(T)$ and $\phi_{2}^{S}(T)$ given
by (\ref{eq:defxS1}) and (\ref{eq:defxS2}) respectively, and the
function $\phi_{3}^{M}(T)$ is
\begin{eqnarray}
\phi_{3}^{M}(T)=&(L-v_1)^2+(L-v_2T)^2+i[(v_1^2-v_2^2)T^2-2L(v_1-v_2)T]T/T^L
\nonumber \\ &-i4\sigma_r^2(\bar E_1-\bar E_2)T+i4\sigma_r^2(\bar
p_1-\bar p_2)L. \label{eq:defxNS3}
\end{eqnarray}
where we have neglected the terms with powers higher than
$\textit{O}(T/T^T)$ and $\textit{O}(T/T^L)$. Now we focus our
attention on the elimination of the time dependence that is
present in the neutrino oscillation probabilities
(\ref{eq:nudpt5b}) and (\ref{eq:edpt5b}). To do it, we take the
average on the time of the expressions (\ref{eq:nudpt5b}) and
(\ref{eq:edpt5b}). With the integration on the time we obtain
\begin{equation}
P_{\nu_e}^{MDWP}(L)=\frac{1}{(2\pi\sigma_r^2)^{1/2}}\frac{1}{(1+\Lambda^2)^2}
\left\{ \Lambda^4\,I_1^M+ I_2^M+ \Lambda^2\aleph\,I_3^M \right\}
\label{eq:nudpt5c},
\end{equation}
\begin{equation}
P_{\nu_\mu}^{MDWP}(L)=\frac{1}{(2\pi\sigma_r^2)^{1/2}}\frac{1}{(1+\Lambda^2)^2}
\left\{ \Lambda^2\,I_1+ \Lambda^2\,I_2- \Lambda^2\aleph\,I_3
\right\} \label{eq:edpt5c},
\end{equation}
where the integrals on the time for the minimum dispersion regime
$I_1^M$, $I_2^M$ and $I_3^M$ are
\begin{align}
I_1^M&=\int \exp[-\lambda_1\phi_{1}^{S}(T)] dT,\\
I_2^M&=\int \exp[-\lambda_2\phi_{2}^{S}(T)] dT,\\
I_3^M&=\int \exp[-\lambda_3\phi_{3}^{M}(T)] dT.\\
\end{align}
The integrals $I_1^M$ and $I_2^M$ can be performed using both
Gaussian integration and the Laplace approximation method, while
the integral $I_3^M$  can be only performed using the Laplace
approximation method. After the time integrations are performed,
we obtain from (\ref{eq:nudpt5c}) and (\ref{eq:edpt5c}) the
following time-integrated neutrino oscillation probabilities
\begin{equation} \label{eq:nudpt5d}
P_{\nu_e}^{MDWP}(L)=\frac{1}{(1+\Lambda^2)^2}\left\{\frac{
\Lambda^4}{v_1}+\frac{1}{v_2}+\Lambda^2\Xi\left(
\frac{2}{v_1^2+v_2^2} \right)^{1/2} f_{3}^{T} \exp\left[i
f_{1}^{T}-f_{2}^{T}\right]\right\},
\end{equation}
\begin{equation} \label{eq:edpt5d}
P_{\nu_\mu}^{MDWP}(L)=\frac{1}{(1+\Lambda^2)^2}\left\{\frac{
\Lambda^2}{v_1}+\frac{\Lambda^2}{v_2}-2\Lambda^2\Xi\left(
\frac{2}{v_1^2+v_2^2} \right)^{1/2} f_{3}^{T} \exp\left[i
f_{1}^{T}-f_{2}^{T}\right]\right\},
\end{equation}
with the functions $f_{1}^{T}$, $f_{2}^{T}$ and $f_{3}^{T}$ given
by
\begin{align}
f_{1}^{T}&=(\bar E_1 - \bar
E_2)\frac{v_1+v_2}{v_1^2+v_2^2}L(1-h_{1}^{T})-(\bar p_1 -\bar p_2)L,\\
f_{2}^{T}&=\frac{(v_1 - v_2)^2}{v_1^2 +
v_2^2}\frac{L^2}{4\sigma_r^2}(1+h_{2}^{T}) + \frac{(\bar
E_1- \bar E_2)^2}{v_1^2 + v_2^2} \sigma_r^2(1-h_{3}^{T}),\\
f_{3}^{T}&=\left(1+\frac{6(v_1^2-v_2^2)(\bar E_1 - \bar E_2)\bar
S}{(v_1^2 +v_2^2)^2}\right)^{-1/2},
\end{align}
where
\begin{align}
h_{1}^{T}&=-\frac{1}{2}\frac{(\bar E_1 - \bar
E_2)(v_1-v_2)}{(v_1 +v_2)(v_1^2 + v_2^2)^2}\left[(v_1^2 +v_2^2)^2 +4v_1 v_2 \right]\bar S,\\
h_{2}^{T}&=-\frac{1}{2}\frac{(\bar E_1 -\bar E_2)(v_1 +v_2)}{(v_1 - v_2)(v_1^2 + v_2^2)},\\
h_{3}^{T}&=-\frac{5}{2}\frac{(\bar E_1 -\bar E_2)(v_1^2
-v_2^2)}{(v_1^2 + v_2^2)^2}\bar S,
\end{align}
In the relativistic limit, using the approximations
(\ref{eq:relap1}), (\ref{eq:relap2}) and (\ref{eq:relap3}), we
obtain from (\ref{eq:nudpt5d}) and (\ref{eq:edpt5d}) that the
time-integrated neutrino oscillation probabilities for the minimum
dispersion regime are
\begin{eqnarray} \label{eq:nudpt5e}
P_{\nu_e}^{MDWP}(L)=1-&\frac{1}{2}\sin^2[2\theta_{12}]\left\{1-
\frac{1}{(1-a_4)^{1/2}}\exp\left[i\,2\pi\frac{L}{L'_{osc}}-\left(\frac{L}{L'_{coh}}\right)^2
\right. \right. \nonumber \\ &\left. \left. - 2\pi^2
\xi^2\left(\frac{\sigma_r(1-a_3)}{L_{osc}}\right)^2\right]\right\},
\end{eqnarray}
\begin{eqnarray} \label{eq:edpt5e}
P_{\nu_\mu}^{MDWP}(L)=&\frac{1}{2}\sin^2[2\theta_{12}]\left\{1-
\frac{1}{(1-a_4)^{1/2}}\exp\left[i\,2\pi\frac{L}{L'_{osc}}-\left(\frac{L}{L'_{coh}}\right)^2
\right. \right. \nonumber \\ &\left. \left.- 2\pi^2
\xi^2\left(\frac{\sigma_r(1-a_3)}{L_{osc}}\right)^2\right]\right\},
\end{eqnarray}
where $L'_{osc}$ and $L'_{coh}$ are written as
\begin{align}
L'_{osc}=&\frac{L_{osc}}{1+a_1},\label{eq:LOscpri}\\
L'_{coh}=&\frac{L_{coh}}{(1+a_2)^{1/2}},\label{eq:LCohpri}
\end{align}
and
\begin{align}
a_1&=\frac{1}{8}\frac{\xi^2 (\Delta m_{12}^2)^2\bar m^2}{\bar
E^6}=
4\xi^2\frac{\sigma_r^2}{L_{coh}^2}\frac{T^T}{T^L},\\
a_2&=5\xi\frac{\bar m^2}{\bar E^2}=5\xi\frac{T^T}{T^L},\\
a_3&=\frac{5}{16}\frac{\xi (\Delta m_{12}^2)^2\bar m^2}{\bar E^6}=
10\xi\frac{\sigma_r^2}{L_{coh}^2}\frac{T^T}{T^L},\\
a_4&=\frac{3}{4}\frac{\xi (\Delta m_{12}^2)^2\bar m^2}{\bar E^6}=
24\xi\frac{\sigma_r^2}{L_{coh}^2}\frac{T^T}{T^L}.
\end{align}
We can observe that the time-integrated oscillation probabilities
(\ref{eq:nudpt5e}) and (\ref{eq:edpt5e}) have the same functional
form that the oscillation probabilities (\ref{eq:nudpt4d}) and
(\ref{eq:edpt4d}) obtained considering the expansion of energy
until second order in the momentum, but now the exponential is
multiplied by a factor that includes $a_4$. But new, due to the
effects originated in dispersion with time on the spreading of the
wave packets, the expressions (\ref{eq:LOscpri}) and
(\ref{eq:LCohpri}) show respectively some changes of the
oscillation length (\ref{eq:LOsc}) and of the coherence length
(\ref{eq:LCoh}). We observe how the oscillation length
(\ref{eq:LOscpri}) and the coherence length (\ref{eq:LCohpri}) are
respectively a little smaller than the ones obtained for the case
in which the energy is expanded until second order in the momentum
(\ref{eq:LOsc}) and (\ref{eq:LCoh}). Due to the effects originated
in dispersion with time, now the maximum number of oscillations is
\begin{equation}\label{eq:Numospri}
N'_{osc}=\frac{L'_{coh}}{L'_{osc}}=\frac{(1+a_1)}{(1+a_2)^{1/2}}N_{osc},
\end{equation}
which implies that it is smaller than the one obtained for the
case in which the energy is expanded until second order in the
momentum (see the expression (\ref{eq:Numos})). However, the
quantities $a_i$, with $i=1,2,3,4$, are very small, so
$L'_{osc}\simeq L_{osc}$, $L'_{coh}\simeq L_{coh}$, $N'_{osc}
\simeq N_{osc}$, $a_3 \simeq 0$ and $a_4 \simeq 0$. In this way,
the time-integrated oscillation probabilities (\ref{eq:nudpt5e})
and (\ref{eq:edpt5e}) can reduce to (\ref{eq:nudpt4d}) and
(\ref{eq:edpt4d}).  Thus, for the minimum dispersion regime we
find that the effects originated in dispersion with time on the
spreading of the wave packets for the time-integrated neutrino
oscillation probabilities can be neglected.

\subsection{Spreading in the transversal dispersion regime}

In the transversal dispersion regime $T^T<T<T^L$ and for the limit
of nearly degenerate masses, the oscillation probabilities
(\ref{eq:nudpt5a}) and (\ref{eq:edpt5a}) can be written as
\begin{eqnarray}
P_{\nu_e}^{TDWP}(T,L)=&\frac{1}{(2\pi\sigma_r^2)^{1/2}}\frac{1}{(1+\Lambda^2)^2}
\left\{ \Lambda^4 F^T(T)\,\exp[-\lambda_1\phi_{1}^{S}(T)]\right.
\nonumber \\ &\left. + F^T(T)\exp[-\lambda_2\phi_{2}^{S}(T)]+
\Lambda^2\aleph F^T(T)\,\exp[-\lambda_3\phi_{3}^{M}(T)] \right\}
\label{eq:nudpt5f},
\end{eqnarray}
\begin{eqnarray}
P_{\nu_\mu}^{TDWP}(T,L)=&\frac{1}{(2\pi\sigma_r^2)^{1/2}}\frac{1}{(1+\Lambda^2)^2}
\left\{ \Lambda^2 F^T(T)\,\exp[-\lambda_1\phi_{1}^{S}(T)]\right.
\nonumber \\ &\left. + \Lambda^2 F^T(T)
\,\exp[-\lambda_2\phi_{2}^{S}(T)]- \Lambda^2\aleph
F^T(T)\,\exp[-\lambda_3\phi_{3}^{M}(T)] \right\}
\label{eq:edpt5f},
\end{eqnarray}
where the function on the time $F^T(T)$ is given by
$F^T(T)=\frac{(T^T)^2}{T^2}$. In the expressions
(\ref{eq:nudpt5f}) and (\ref{eq:edpt5f}), we have neglected the
terms with powers higher than $\textit{O}(T/T^L)$ and the
functions $\phi_{1}^{S}(T)$ and $\phi_{2}^{S}(T)$ are given by
(\ref{eq:defxS1}) and (\ref{eq:defxS2}) respectively, while the
function $\phi_{3}^{M}(T)$ is given by (\ref{eq:defxNS3}). The
three integrals that appear in the expressions (\ref{eq:nudpt5f})
and (\ref{eq:edpt5f}) can be only performed using the Laplace
approximation method. After the time integrations are performed,
we obtain from (\ref{eq:nudpt5f}) and (\ref{eq:edpt5f}) the
following time-integrated neutrino oscillation probabilities
\begin{equation} \label{eq:nudpt5g}
P_{\nu_e}^{TDWP}(L)=\frac{(T^T)^2}{L^2}P_{\nu_e}^{MDWP}(L)=
\frac{(T^T)^2}{L^2}P_{\nu_e}^{SWP}(L),
\end{equation}
\begin{equation} \label{eq:edpt5g}
P_{\nu_\mu}^{TDWP}(L)=\frac{(T^T)^2}{L^2}P_{\nu_\mu}^{MDWP}(L)=
\frac{(T^T)^2}{L^2}P_{\nu_\mu}^{SWP}(L),
\end{equation}
where the standard time-integrated neutrino oscillation
probabilities $P_{\nu_e}^{SWP}(L)$ and $P_{\nu_\mu}^{SWP}(L)$ are
given respectively by (\ref{eq:nudpt4d}) and (\ref{eq:edpt4d}),
and the transversal dispersion time by $T^T=2 \bar E \sigma_r^2$.
We observe for this case that the standard time-integrated
neutrino oscillation probabilities $P_{\nu_e}^{SWP}(L)$ and
$P_{\nu_\mu}^{SWP}(L)$ are suppressed by a factor $(T^T)^2/L^2$.
This result is in agreement with the showed by Naumov
\cite{Naumov2013}, whom has recently obtained that the integral
over time of both the flux and probability densities are
asymptotically proportional to the factor $1/L^2$, when he
considered a theory of wave packets in which the energy that
appears in the wave packets is expanded until third order in the
momentum. This author has demonstrated that the origin of the
factor $1/L^2$ for quantum objects is their dispersion with time
\cite{Naumov2013}.

\subsection{Spreading in the longitudinal dispersion regime}

In the longitudinal dispersion regime $T>T^L$ and for the limit of
nearly degenerate masses, the oscillation probabilities
(\ref{eq:nudpt5a}) and (\ref{eq:edpt5a}) can be written as
\begin{eqnarray}
P_{\nu_e}^{LDWP}(T,L)=&\frac{1}{(2\pi\sigma_r^2)^{1/2}}\frac{1}{(1+\Lambda^2)^2}
\left\{ \Lambda^4 F^L(T)\,\exp[-\lambda_1\phi_{1}^{D}(T)]\right.
\nonumber \\ &\left. + F^L(T)\exp[-\lambda_2\phi_{2}^{D}(T)]+
\Lambda^2\aleph F^L(T)\,\exp[-\lambda_3\phi_{3}^{D}(T)] \right\}
\label{eq:nudpt5h},
\end{eqnarray}
\begin{eqnarray}
P_{\nu_\mu}^{LDWP}(T,L)=&\frac{1}{(2\pi\sigma_r^2)^{1/2}}\frac{1}{(1+\Lambda^2)^2}
\left\{ \Lambda^2 F^L(T)\,\exp[-\lambda_1\phi_{1}^{D}(T)]\right.
\nonumber \\ &\left. + \Lambda^2 F^L(T)
\,\exp[-\lambda_2\phi_{2}^{D}(T)]- \Lambda^2\aleph
F^L(T)\,\exp[-\lambda_3\phi_{3}^{D}(T)] \right\}
\label{eq:edpt5i},
\end{eqnarray}
where the function on the time $F^L(T)$ is given by
$F^L(T)=\frac{(T^T)^2T^L}{T^3}$. In the expressions
(\ref{eq:nudpt5h}) and (\ref{eq:edpt5i}), the functions
$\phi_{1}^{D}(T)$, $\phi_{2}^{D}(T)$ and $\phi_{3}^{D}(T)$ are
given by (\ref{eq:phi1}), (\ref{eq:phi2}) and (\ref{eq:phi3})
respectively. For this case, also the three integrals that appear
in the expressions (\ref{eq:nudpt5h}) and (\ref{eq:edpt5i}) can be
only performed using the Laplace approximation method. After the
time integrations are performed, we obtain from (\ref{eq:nudpt5h})
and (\ref{eq:edpt5i}) the following time-integrated neutrino
oscillation probabilities
\begin{equation} \label{eq:nudpt5j}
P_{\nu_e}^{LDWP}(L)=\frac{(T^T)^2}{L^2}P_{\nu_e}^{MDWP}(L)=
\frac{(T^T)^2}{L^2}P_{\nu_e}^{SWP}(L),
\end{equation}
\begin{equation} \label{eq:edpt5k}
P_{\nu_\mu}^{LDWP}(L)=\frac{(T^T)^2}{L^2}P_{\nu_\mu}^{MDWP}(L)=
\frac{(T^T)^2}{L^2}P_{\nu_\mu}^{SWP}(L),
\end{equation}
where the standard time-integrated neutrino oscillation
probabilities $P_{\nu_e}^{SWP}(L)$ and $P_{\nu_\mu}^{SWP}(L)$ are
given respectively by (\ref{eq:nudpt4d}) and (\ref{eq:edpt4d}). We
observe also for this case that the standard time-integrated
neutrino oscillation probabilities $P_{\nu_e}^{SWP}(L)$ and
$P_{\nu_\mu}^{SWP}(L)$ are suppressed by a factor $(T^T)^2/L^2$.

\section{Conclusions}

We have studied the effects originated in dispersion with time on
spreading of wave packets for the time-integrated two-flavor
neutrino oscillation probabilities in vacuum. We have calculated
the time-integrated two-flavor neutrino oscillation probabilities
in the context of a wave packet extension of the quantum field
theory treatment that we previously developed for the case in
which neutrino mass eigenstates were described by plane waves. In
the treatment that we have presented here, neutrino flavor states
have been considered as superpositions of neutrino mass
eigenstates described by localized wave packets.

By methodological reasons, we have initially studied the effects
of the spreading of the wave packets by considering the expansion
of the energy until second order in the momentum that leads to the
standard time-integrated neutrino oscillation probabilities. After
this, we have studied the effects originated by dispersion in time
on spreading of wave packets for the time-integrated two-flavor
neutrino oscillations by considering the expansion of the energy
until third order in the momentum. We have observed that the
standard time-integrated neutrino oscillation probabilities are
suppressed by a factor $1/L^2$ for the transversal and
longitudinal dispersion regimes, where $L$ is the distance between
the neutrino source and the detector. The existence of this kind
of suppression for the standard time-integrated neutrino
oscillation probabilities might be proved in reactor neutrino
oscillation experiments with beams very narrow in time or
experiments at short enough distances \cite{Naumov2013}.

\section*{Acknowledgments}

Y. F. P\'{e}rez thanks for the financial support from the
brazilian supporting agencies Funda\c{c}\~{a}o de Amparo \`{a}
Pesquisa do Estado de S\~{a}o Paulo (FAPESP) and Conselho Nacional
de Desenvolvimento Cient\'{\i}fico e Tecnol\'{o}gico (CNPq). C. J.
Quimbay thanks DIB for the financial support received through the
research project "Propiedades electromagn\'{e}ticas y de
oscilaci\'{o}n de neutrinos de Majorana y de Dirac". C. J. Quimbay
would like thank to Maurizio De Sanctis for help in the manuscript
preparation.

\end{document}